\documentclass{my_roaj}
\usepackage[dvips]{graphicx}

\usepackage[latin1]{inputenc}
\usepackage[T1]{fontenc}
\usepackage[french]{babel}

\usepackage{amsmath,amssymb}

\Volume{Vol. 13}
\Issue{1}
\Pagespan{31}{}
\Yearpublication{2003}
\Yearsubmission{2002}

\begin{document}

\title{Contribution à l'etude des binaires des types F, G, K, M\\
IX. HD 191588, nouvelle binaire spectroscopique à raies
simples de type RS Cvn, systeme triple.} 

\titlerunning{HD 191588, système triple}

\author{
R.F. Griffin\inst{1}
N. Ginestet\inst{2}
et
J.-M. Carquillat\inst{2}
} 
\institute{
The Observatories, Madingley Road, Cambridge CB3 0HA, England
\and
Observatoire Midi-Pyrénées, UMR 5572, 14, avenue Edouard Belin, 
F-31400 Toulouse, France
}
\authorrunning{Roger GRIFFIN, Nicole GINESTET et Jean-Michel CARQUILLAT}

\received{September 2002} 
\accepted{December 2002}


\abstract{%
\\
\\
\centerline{\hglue -2cm \bf CONTRIBUTION TO THE STUDY OF F, G, K, M BINARIES}\\
\centerline{\hglue -2cm \bf IX. HD 191588, A NEW RS CVN-TYPE SINGLE-LINED}\\
\centerline{\hglue -2cm \bf SPECTROSCOPIC  BINARY, A TRIPLE SYSTEM}\\
\\
\\
Abstract: 
An accident of misidentification has brought to light 
the interesting system HD 191588, a new RS CVn-type spectroscopic
binary. 
A radial-velocity study of the primary star, the only seen component, 
carried out at the Observatoire de Haute-Provence with the Coravel 
instrument and subsequently at the Cambridge Observatories with a
similar one, 
reveals two orbital motions: a short-period orbit (60 days) 
and a long-period one (about 4.5 years), so this star is a triple
system.\\
The following orbital elements are obtained:
(1) for the long-period orbit $P = 1667 \pm17$~days, $T = 50901 \pm
67$~MJD, 
$\Gamma = +2.09 \pm0.07$ km s$^{-1}$, $K = 2.51 \pm0.13$ km s$^{-1}$, 
$e = 0.18 \pm0.04$, $\omega = 228^{\circ} \pm 14^{\circ}$, 
$a_1 \sin i = 56.7 \pm 3.0$~Gm, $f(m) = 0.0026 \pm0.0004$~M$_\odot$, 
and (2) for the short-period orbit $P = 60.0269 \pm0.0016$~days, 
$T = 50482.6 \pm3.3$~MJD, $\gamma$ is var., $K = 24.03 \pm 0.09$~km
s$^{-1}$, 
$e = 0.012 \pm0.004$, $\omega = 233^{\circ} \pm19^{\circ}$, 
$a_1 \sin i = 19.83 \pm 0.07$~Gm, $f(m) = 0.0865 \pm0.0009$~M$_\odot$.\\ 
From near-infrared observations we refine the classification of the
primary component and we found a spectral type of K2.5 III, and a spectrum
obtained in the blue--near-UV spectral region reveals strong H and K emission
lines of Ca~II. The unseen secondary should be a solar-type star (F or G V); 
the minimum mass of the third body is that of a dwarf M star.
Probably, the primary component rotates in synchronism with the orbital
motion in the inner orbit; a model, based upon that hypothesis, is
proposed for the system, and finally the connection of the inner binary 
to the long-period RS CVn group (Hall 1976) is discussed.
} 

\maketitle

\begin{figure*}
\vglue-2.7cm
\hglue-3cm
\includegraphics*{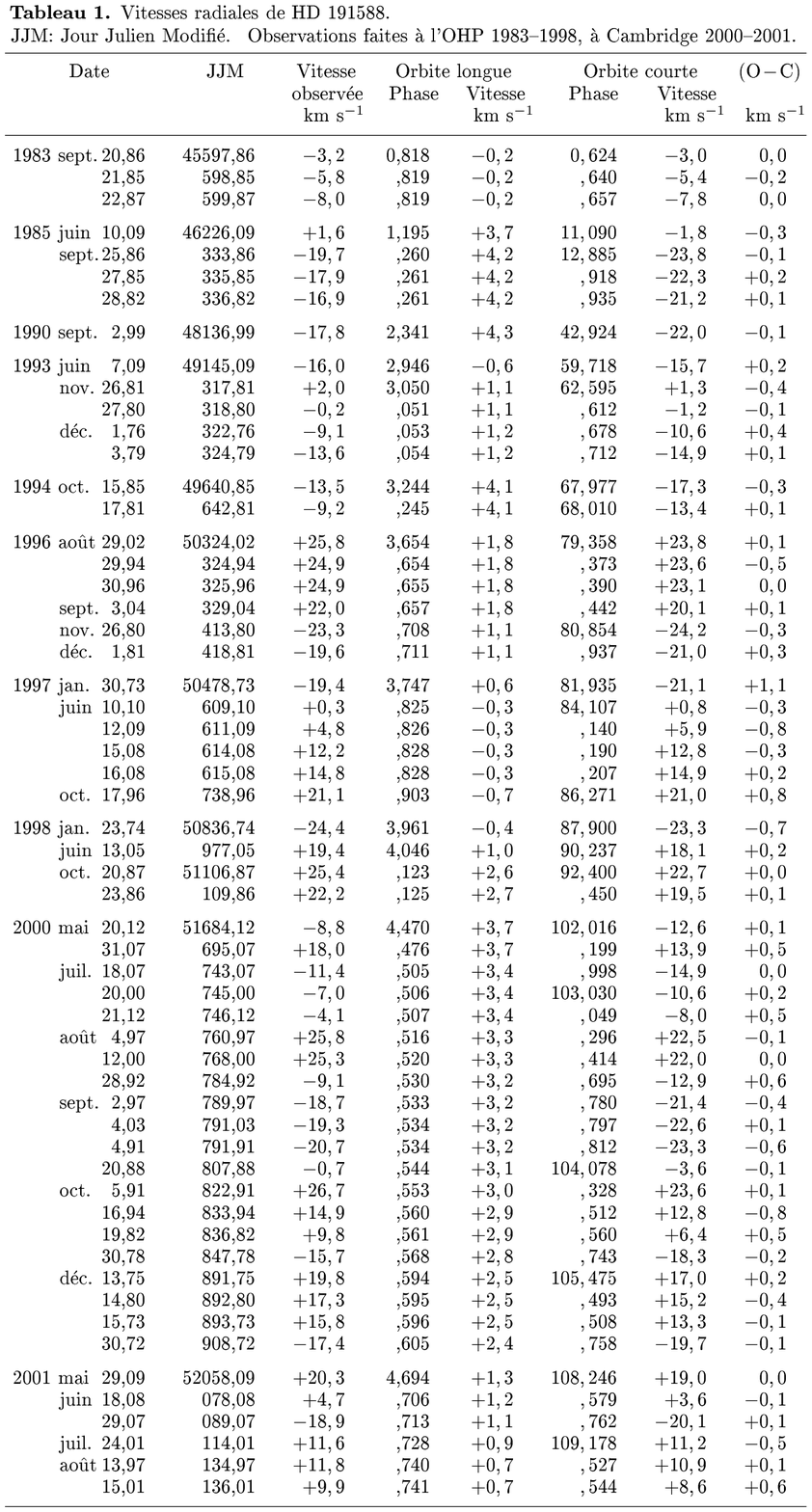}
\end{figure*}

\section{INTRODUCTION}

Cet article est le neuvième d'une série consacrée à l'étude des
vitesses radiales d'étoiles des derniers types spectraux (F à M)
signalées comme binaires spectroscopiques mais dont les éléments
orbitaux n'étaient pas connus. 
Ce sont cependant des circonstances tout à fait fortuites qui nous 
ont amenés à étudier l'intéressant système HD 191588 (HIP 99301). 
En effet, lorsque l'équipe de Toulouse a entamé, au début des années 1980, 
ses observations systématiques en vitesse radiale d'un échantillon d'étoiles 
à spectre composite à l'aide de l'instrument CORAVEL de l'observatoire 
de Haute-Provence (OHP), l'étoile HDE 227984 (étoile 507 de la liste de Hynek, 
1938), très proche en position de HD 191588, figurait à son programme. 
Une confusion (erreur de pointage) a donc été faite, et pendant 
un temps HD 191588 a été observée 
en croyant qu'il s'agissait de HDE 227984. Les observateurs toulousains 
avaient certes remarqué que l'objet en question paraissait nettement plus 
brillant que la magnitude $m_{\mathrm{ptm}} = 10$ 
reportée dans la liste de Hynek, mais consultant d'abord la base de données 
SIMBAD (CDS de Strasbourg), ils avaient été, dans un premier temps, rassurés: 
cette base donnait HDE 227984 avec la magnitude $m_v = 8,3$, 
ce qui correspondait bien à ce qu'ils observaient. 
Ce qu'ils ne savaient pas, c'est que la même erreur avait été faite au CDS 
de Strasbourg! Soulignons néanmoins qu'à présent cette méprise a été corrigée 
dans SIMBAD où HDE 227984 est ``redevenue'' l'étoile faible répertoriée 
par Hynek. 

Dans cet article nous montrons que HD 191588 est en fait un système triple 
spectroscopique dont la composante principale, la seule visible, est 
une géante froide présentant de fortes raies H et K en émission dans 
son spectre.

Hormis les données figurant dans le Henry Draper Catalogue 
($m_\mathrm{ptm} = 8,4$; $m_\mathrm{ptg} = 9,5$; Sp: K2), 
celles dont nous disposons sur HD 191588 ont été obtenues au cours 
d'investigations effectuées à l'OHP sur plusieurs champs galactiques, 
et dans la constellation du Cygne pour ce qui concerne cette étoile. 
Ces investigations mettaient en \oe uvre conjointement des observations 
en photométrie photoélectrique, avec un matériel réalisé à l'Observatoire 
de Toulouse, et des observations spectrographiques avec 
le ``petit prisme objectif'' de Fehrenbach. Les données en photométrie 
photoélectrique sont: $V = 8,32$ et $B - V = 1,17$ (Bouigue et al. 1963), 
cette valeur de l'indice de couleur étant reportée dans le catalogue 
Hipparcos (ESA, 1997). \textsl{Mais dans la publication de Bouigue et al., 
ces paramètres sont attribués à tort à l'étoile HDE  227984, et 
c'est sans doute de là que l'erreur s'est répercutée jusqu'au CDS.}

A partir des observations au prisme objectif, Duflot et al. (1958) donnent, 
avec 4 clichés, une vitesse radiale de $-15$~km s$^{-1}$, à laquelle ils 
attribuent la qualité B (2,5 km s$^{-1}$ $< \varepsilon <$ 5 km s$^{-1}$); 
ces observations leur permettent aussi de donner la classification G8 III, 
qui sera retenue dans la base SIMBAD. Notons qu'aucune remarque n'est 
formulée par ces observateurs au sujet d'une éventuelle variabilité 
de la vitesse radiale de HD 191588.

Ces données sur les champs galactiques étudiés à l'OHP ont ensuite été 
utilisées par Boulon (1963) pour son travail de thèse. A cette occasion, 
Boulon révise la classification de HD 191588, qui devient K0 II. Mais la 
classe de luminosité II apparaît maintenant irréaliste car elle impliquerait 
une distance de quelque 1 kpc alors que celle déduite de la parallaxe 
Hipparcos est comprise entre seulement 210 et 350 pc.

Afin de procéder nous-mêmes à une classification spectrale de cette étoile, 
et pensant encore qu'il s'agissait d'un spectre composite, nous l'avons 
observée avec le spectrographe Aurélie de l'OHP dans deux domaines spectraux 
distincts: (1) le proche infrarouge (8400--8800 \AA), 
dispersion de 33 \AA/mm, et (2) le bleu--proche UV (3800--4230 \AA), 
dispersion 16 \AA/mm. Le spectre obtenu dans le proche IR nous a conduits 
à la classification K2,5 III Ca- (Ginestet et al. 1999; 
N.B.: dans la Table 2 de cet article, lire HD 191588 à la place de HDE 227984),
tandis que celui pris dans le bleu mettait en évidence une forte émission 
des raies H et K de Ca II (Fig. 1). Cette émission n'apparaissait pas 
directement sur les raies, en absorption, du triplet de Ca II dans 
le proche IR, mais elle devait certainement les combler partiellement 
et en diminuer l'intensité (Linsky et al. 1979).
\begin{figure*}
\includegraphics*[width=2.65cm,angle=-90]{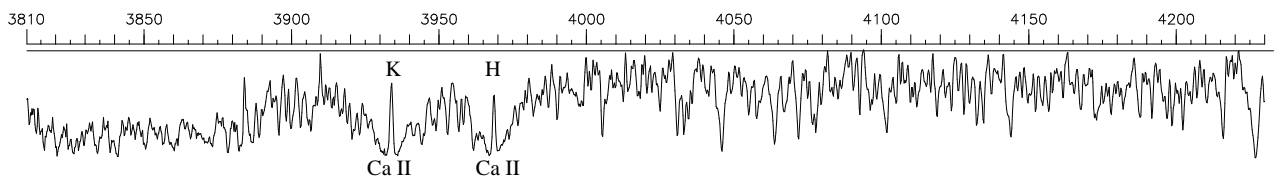}
\caption{%
Spectre de HD 191588 dans la région spectrale 3800--4230 \AA\ montrant 
la présence des raies H et K de Ca II en émission.} 
\end{figure*}

\section{Obtention des vitesses radiales et des éléments orbitaux}

HD 191588 a été observée de septembre 1983 à octobre 1998 avec 
le spectrovélocimètre CORAVEL (Baranne et al. 1979) monté au foyer 
cassegrain du télescope suisse de 1~m de l'OHP et en 2000--2001 avec 
un instrument assez similaire, monté au foyer coudé du télescope de 91 cm 
de l'Observatoire de Cambridge (UK). Trente et une vitesses radiales (VR) 
ont été obtenues à l'OHP et vingt six à Cambridge (Tableau 1). 
Ces VR sont ramenées au système actuellement adopté pour la base de 
données CORAVEL-OHP (Udry et al. 1999); dans ce but, les observations 
faites à Cambridge et réduites initialement par la méthode habituelle 
(Griffin 1969) ont reçu un ajustement de $-1$~km~s$^{-1}$.
 
Une première analyse mettait en évidence une orbite quasi circulaire 
avec une période voisine de 60 jours, mais on remarquait sur la courbe 
de VR une dispersion des points beaucoup trop forte comparée à 
la précision des mesures. Cette dispersion anormale des points pouvait 
provenir de l'action perturbatrice d'un troisième corps qui entraîne 
le couple suivant un mouvement képlérien avec une période de 4 ans et demi. 
Nous avons déterminé simultanément les deux orbites, 
à longue et à courte périodes, au moyen d'un programme de calcul 
écrit initialement par A.P. Cornell et R.F.G.; dans le cas présent, 
l'orbite à longue période (que nous désignerons par ``orbite longue'') 
est résolue par la méthode de Lehmann-Filhes (1894) 
et celle à courte période (que nous désignerons par ``orbite courte''), 
qui est presque circulaire, est calculée par la méthode de Sterne (1941). 
Le principe de la détermination des éléments est que la valeur de $\gamma$, 
vitesse du centre de masses pour l'orbite courte, est défini par 
les éléments de l'orbite longue, la vitesse systématique de 
l'ensemble du système étant représentée par $\Gamma$; 
chaque vitesse radiale pour l'orbite courte est corrigée de la 
variation de $\gamma$ en fonction du temps, la valeur de $\gamma$ 
est donc ramenée à zéro comme le fait apparaître le graphique 
correspondant à cette orbite. Les courbes de VR pour les deux orbites 
sont représentées aux Figs 2a et 2b; à la Fig. 2a la variation due 
à l'orbite courte a été déduite de sorte que seule demeure la variation 
imputable à l'orbite longue, tandis qu'à la Fig. 2b c'est le mouvement 
provenant de l'orbite longue qui a été déduit afin d'isoler la variation 
de VR due à l'orbite courte. Au Tableau 1 nous donnons les VR observées 
ainsi que les phases et les vitesses calculées correspondant aux deux 
mouvements orbitaux; par suite de la méthode de calcul utilisée, 
le résidu $O - C$ pour chaque observation sera le même, que l'on considère 
l'orbite longue ou l'orbite courte. Les variances relatives 
aux VR observées tant à l'OHP qu'à Cambridge apparaissent quasi identiques, 
si bien que les 57 mesures ont été affectées du même poids pour la 
détermination des orbites. Les éléments orbitaux obtenus, avec leurs 
erreurs standards, sont présentés en Tableau 2.

\begin{figure*}
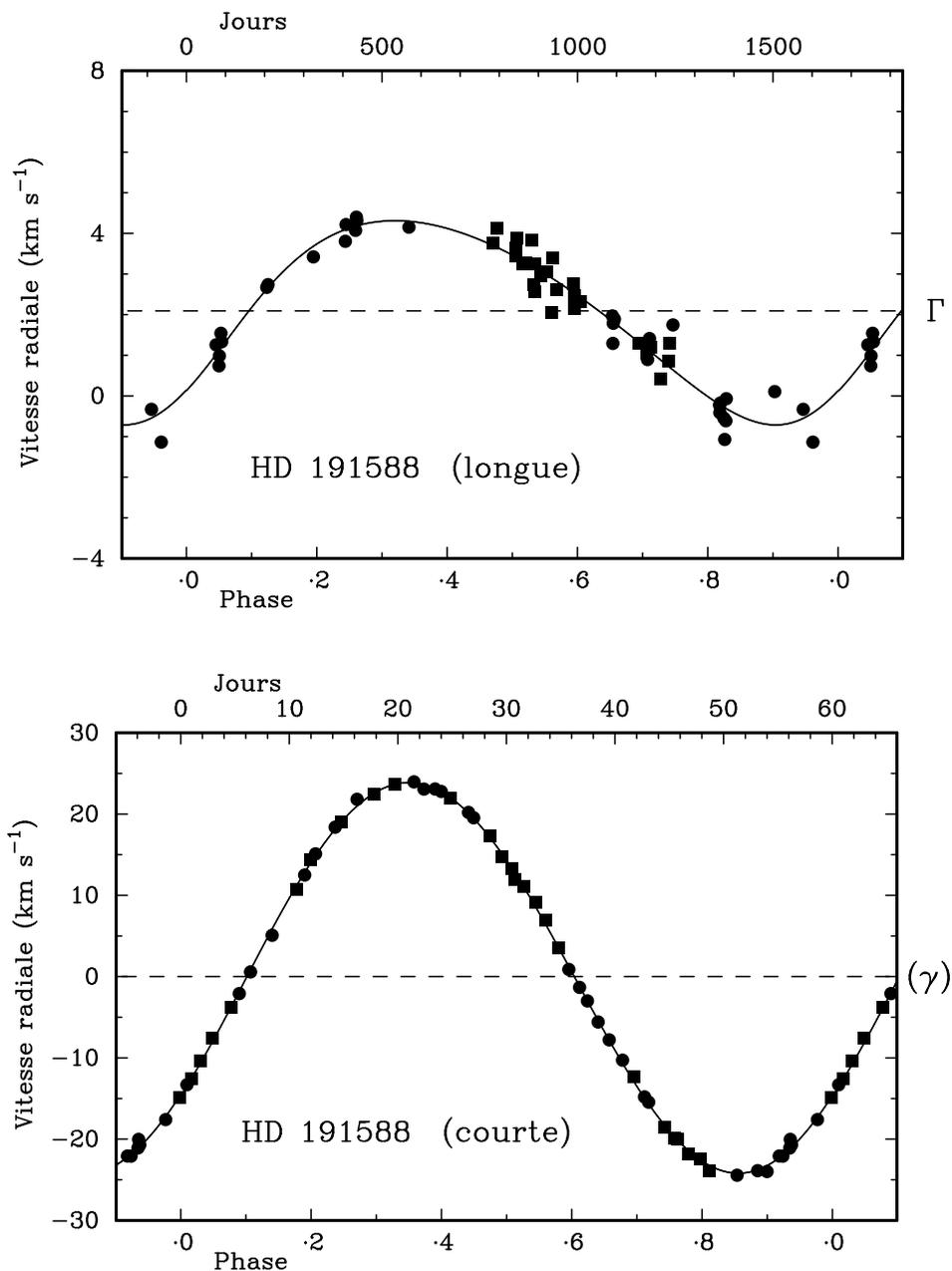

\centerline{
\includegraphics*[width=8cm,angle=-90]{HD191588Fig2a.ps}
}
\vskip 8mm
\centerline{
\includegraphics*[width=8cm,angle=-90]{HD191588Fig2b.ps}
}
\caption{Vitesses radiales observées et courbes de VR calculées 
de HD 191588 en fonction de la phase pour l'orbite à longue période 
(en haut, Fig. 2a) 
et pour celle à courte période (en bas, Fig. 2b). Les cercles 
correspondent aux observations effectuées à l'OHP, les carrés à celles 
faites à Cambridge.}
\end{figure*}

\addtocounter{table}{+1}
\begin{table}
\caption{Eléments orbitaux du système triple spectroscopique HD 191558.
$T$ est l'époque du passage au périastre, 
$T_0$ l'époque du passage au n\oe ud ascendant.} 
\begin{tabular}{ccc}
\hline
 & & \\
Elément & Orbite longue & Orbite courte \\
 & & \\
\hline
 & & \\
$P$  [jours] & 1667 $\pm$ 17 & 60,0269 $\pm$ 0,0016 \\
$T_0$   [JJM] & & 50443,76 $\pm$ 0,04 \\
$T$    [JJM] & 50901 $\pm$ 67 & 50482,6 $\pm$ 3,3 \\
$\Gamma$ [km s$^{-1}$] & $+$2,09 $\pm$ 0,07 & \\
$K$    (km s$^{-1}$] & 2,51 $\pm$ 0,13 & 24,03 $\pm$ 0,09 \\
$e$ & 0,18 $\pm$ 0,04 & 0,0116 $\pm$ 0,0038 \\
$\omega$  [degrés] & 228 $\pm$ 14 & 233 $\pm$ 19 \\
$a_1 \sin i$  [Gm] & 56,7 $\pm$ 3,0 & 19,83 $\pm$ 0,07 \\
$f(m)$   [M$_\odot$] &  0,0026 $\pm$ 0,0004 & 0,0865 $\pm$ 0,0009 \\
 & & \\
\hline
\end{tabular}
\end{table}

Le but initial des observations effectuées à Cambridge était de combler 
les manques d'observations à certaines phases de l'orbite à courte période. 
De telles discontinuités dans la courbe de VR observée sont en effet 
difficiles à combler avec des missions d'observations d'une durée d'environ 
une semaine tous les trimestres, voire tous les semestres comme c'était 
le cas à l'OHP, surtout quand la période orbitale de la binaire observée 
est nettement supérieure à la durée d'une ``mission''. Par chance, 
les observations faites à Cambridge  durant l'année 2000 ont permis également 
de pourvoir un intervalle de phases jusqu'ici dégarni sur la courbe de VR 
de l'orbite longue, et d'assurer ainsi la validité de cette dernière.

L'excentricité de l'orbite à courte période est vraiment faible 
--- à peine trois fois son erreur standard --- et on peut se demander 
si elle est significativement différente de zéro. Il est cependant connu 
(Mazeh, 1990) que l'existence d'un troisième corps dans un système 
dynamique peut créer ou maintenir une excentricité dans un système 
qui autrement serait, selon toute vraisemblance, circularisé; 
en conséquence, la présence d'une faible excentricité pour l'orbite 
de période 60 jours de HD 191588 n'est pas surprenante même si la plupart 
des binaires comportant une étoile géante ont des orbites véritablement 
circulaires pour des périodes aussi courtes. La question peut aussi être 
abordée en utilisant l'un des tests préconisés par Bassett (1978) dans 
lequel on compare, à partir des mêmes données en VR, la somme des carrés 
des résidus, $\sum (O - C)^2$, pour une orbite calculée en fixant 
à zéro l'excentricité et la même quantité pour une orbite calculée 
en ne fixant pas a priori la valeur de cet élément. Dans le cas présent,
la somme $\sum$ pour la solution représentée par les éléments donnés 
ci-dessus, où $e$ n'a par été fixée, est 7,74 (km s$^{-1})^2$, 
tandis que lorsque $e$ est fixée à zéro pour l'orbite à courte période 
la somme $\sum$ atteint 8,98 (km s$^{-1})^2$. 
Dans la solution complète nous avons ajusté en tout 11 éléments orbitaux 
aux 57 observations, ce qui laisse donc 46 degrés de liberté, de sorte 
que chaque degré de liberté est associé à une variance de: 
7,74 / 46 = 0,168 (km s$^{-1})^2$. D'un autre côté, en éliminant $e$ 
et $\omega$ du calcul, nous gagnons deux degrés de liberté supplémentaires 
au prix d'une élévation de $8,98 - 7,74 = 1,24$ (km s$^{-1})^2$ de 
la somme $\sum$, donc 0,62 (km s$^{-1})^2$ pour chacun d'eux, soit, 
selon la notation de Bassett, un facteur $T_2$ = 0,62 / 0,168 = 3,69. 
La signification de ce facteur peut être précisée en se référant aux 
tables (Lindley \& Miller, 1953) de rapport de variance, $F$, 
avec 2 et 46 degrés de liberté: le test apparaît significatif au niveau 
de 5\% ($F_{2,46} (5\%) \approx 3,20$) mais pas à 2.5\% (4,00). Nous en 
concluons que l'excentricité est très probablement significative, 
bien que cela ne soit pas absolument certain.

\section{Discussion}

Le satellite Hipparcos a observé HD 191588 et nous disposons donc 
d'une estimation de sa parallaxe, soit $\pi = 3,79 \pm 0,93$~mas (ESA, 1997), 
ce qui place cet objet entre 210 et 350 pc.

Compte tenu de ses coordonnées galactiques ($72^{\circ} 13$, $+0^{\circ} 89$), 
l'excès de couleur devrait être voisin de 0,05 mag. (Lucke, 1978), 
et l'absorption dans le visible de 0,15 mag.; avec la magnitude $V = 8,22$ 
également déduite des observations du satellite 
(Bouigue et al. donnaient la valeur 8,32 mais cette différence pourrait 
résulter d'une réelle variabilité détectée par Hipparcos), cela conduit 
à la magnitude absolue visuelle globale du système $M_V = 1,0 \pm 0,6$ 
et à un indice de couleur $B - V$ ``dérougi'' de 1,12 mag.

\subsection{Séparation des composantes; natures présumées 
de la secondaire et du troisième corps}

En écrivant la fonction de masse sous la forme 
\begin{equation}
f(m) = M_1 \sin^3 i \; \dfrac{\mu^3 }{(1 + \mu)^2}, 
\end{equation}
où $\mu = M_2 / M_1$ est le rapport des masses de la secondaire et 
de la primaire du système à courte période, nous pouvons calculer 
des valeurs minimums pour $\mu$, $M_2$, et $M_3$ (masse du troisième corps) 
suivant les valeurs présumées de la masse $M_1$ de la primaire. 
Pour les étoiles géantes rouges, et contrairement à celles de la 
séquence principale, les masses sont encore mal connues et il est 
admis qu'elles peuvent varier de manière importante suivant l'âge 
de ces étoiles. En conséquence, nous avons effectué des tests 
en attribuant à $M_1$ les valeurs 1,5, 2, et 2,5 M$_\odot$. 
Pour chaque test, nous avons calculé les séparations $a_i$ et $a_e$ 
des composantes du système intérieur (le plus serré) et du système 
extérieur (troisième corps) avec $i = 90^{\circ}$, mais nous avons 
déjà montré (Carquillat et al. 1982) que les valeurs obtenues 
étaient peu sensibles à celle du paramètre $i$. 
Les résultats figurent en Tableau 3.

\begin{table}
\caption{Masses minimales et séparations de la secondaire et 
du troisième corps pour différentes valeurs de la masse de la primaire}
\centerline{%
\begin{tabular}{ccccccc}
\hline
& & & & & & \\
$M_1$ & $\mu_\mathrm{min}$ & $M_{2\, \mathrm{min}}$  
&  $a_i$ & $M_{3\, \mathrm{min}}$ & $a_e$ & $a_e / a_i$ \\
(M$_\odot$)  & & (M$_\odot$) &  (ua) & (M$_\odot$) & (ua) & \\
& & & & & & \\
\hline
& & & & & & \\
1,5 & 0,51 & 0,765 & 0,392 & 0,25 & 3,741 & 9,54 \\
2,0 & 0,45 & 0,900 & 0,426 & 0,30 & 4,050 & 9,51 \\
2,5 & 0,41 & 1,025 & 0,455 & 0,34 & 4,314 & 9,48 \\
& & & & & & \\
\hline
\end{tabular}
}
\end{table}

Une autre contrainte peut être appliquée en faisant intervenir 
le fait que seule la composante principale est visible, ce qui 
implique entre l'étoile primaire et les autres composantes
une différence supérieure à environ deux magnitudes; d'autre part 
la secondaire, ou le troisième corps, ne peuvent être une étoile 
chaude (type A ou plus chaud) car, dans ce cas, le spectre 
présenterait un aspect composite reconnaissable dans le proche UV. 
Par conséquent, compte tenu des éléments ci-dessus, la secondaire 
devrait être une étoile de type solaire plus chaude que la 
primaire (F ou G V suivant la masse de la primaire). Quant au 
troisième corps, sa masse minimum est celle d'une naine froide 
de type M (Schmidt-Kaler, 1982). Nous présentons plus loin dans 
cette analyse un modèle possible faisant intervenir l'ensemble 
des données observationnelles. 
Enfin, le rapport des demi-grands axes de l'orbite longue et de 
l'orbite courte est voisin de 10, et ce système triple peut 
donc être qualifié de hiérarchisé.

\subsection{Synchronisme rotation--révolution}

Pour la primaire, seule composante observable, l'analyse du profil 
des traces de corrélation (méthode décrite dans Benz \& Mayor 1981) 
a permis d'estimer la vitesse de rotation projetée $v \sin i$ 
de HD 191588 à $11,2 \pm 0,2$ km s$^{-1}$ pour les observations 
effectuées à l'OHP et à $12,1 \pm 0,2$ km s$^{-1}$ pour les 
observations effectuées à Cambridge. En fait ces deux valeurs 
n'apparaissent pas définitivement incompatibles si l'on considère que 
l'erreur externe minimale sur les $v \sin i$ est, pratiquement, 
de 1 km s$^{-1}$; en conséquence nous adopterons 
la valeur $v \sin i = 11,6 \pm 1$ km s$^{-1}$.  

Afin de voir s'il pouvait y avoir synchronisme, pour le système intérieur, 
entre rotation axiale de la primaire et mouvement orbital, 
nous avons pratiqué le test de Kitamura \& Kondo (1978). Ce test, 
rappelons-le, consiste à calculer le rayon de la composante 
considérée en supposant que le synchronisme est effectif; il 
implique l'incontournable hypothèse de coplanarité entre 
les plans orbital et équatorial de la composante considérée. 
Si le rayon trouvé est compatible avec son type spectral, 
on peut conclure que le synchronisme est vraisemblable.

En effet, la vitesse équatoriale linéaire de synchronisme 
est liée au rayon de l'étoile et à la période orbitale par la relation: 
$v_\mathrm{syn} = 50,6 R / P$  ($R$ en rayons solaires, $P$ en jours).
Ici, nous n'avons connaissance que de la projection $v sin i$ de 
cette vitesse, mais si nous supposons $35^{\circ} \leq i \leq 90^{\circ}$ 
(condition approximativement requise pour que l'on ait $\mu < 1$), 
cette inégalité et la relation précédente entraînent 
14 R$_\odot$ $\leq R \leq$ 24 R$_\odot$. 
Or, en interpolant les tables de Schmidt-Kaler (1982), 
le rayon moyen d'une étoile de type K2 III est voisin de 20 R$_\odot$ 
et cette valeur apparaît donc compatible avec l'inégalité précédente.

\subsection{Un modèle pour le système à courte période}
 
Nous proposons ci-dessous pour le système un ``modèle synchrone'' 
qui satisfait aux contraintes observationnelles. En effet, 
l'hypothèse du synchronisme, avec  $R_1 = 20$~R$_\odot$, 
implique $i = 43^{\circ}$. En admettant, par exemple, 
$M_1 = 1,75$~M$_\odot$, l'expression de $f(m)$ donne pour masse 
de la secondaire $M_2 = 1,4$~M$_\odot$, ce qui correspond 
au type spectral F5 V (Schmidt-Kaler, 1982). Les paramètres de ce 
modèle, donnés en Tableau 4, ont été obtenus de façon à 
satisfaire au mieux les valeurs de $M_V$ et $B - V$ pour le système 
(en négligeant la contribution du troisième corps). 
Pour effectuer cet ajustement, nous nous sommes basés sur 
les calibrations de Schmidt-Kaler (1982) pour les magnitudes et 
les indices de couleur en fonction des types spectraux MK. Cependant, 
pour la magnitude absolue de la primaire, $M_{V\,1}$, nous avons 
dû adopter la valeur 0,8 au lieu 0,4 (valeur dans la table 
de Schmidt-Kaler pour une étoile K2 III) afin de satisfaire 
aux contraintes. Donc, dans ce modèle, la composante primaire 
serait légèrement moins lumineuse qu'une étoile géante ``normale'' 
de ce type spectral.  
 
\begin{table*}
\caption{Modèle pour le système à courte période 
dans l'hypothèse du synchronisme rotation--révolution, 
et avec $M_1 = 1,75$ M$_\odot$}
\centerline{
\begin{tabular}{cccccccc}
\hline
& & & & & & & \\
Paramètre & Sp$_1$ & Sp$_2$ & $M_{V \,1}$ & $\Delta m_V$ &
$\mu$ & $M_V$ & $B- V$ \\ 
& & & & & & & \\
\hline
& & & & & & & \\
Modèle & K2,5 III & F5 V & 0,8 & 2,7 & 0,8 & 0,7 & 1,12 \\
Observé & K2,5 III e & & & $>$ 2,0 & & $1,0 \pm 0,6$ & 1,12 \\
& & & & & & & \\
\hline
\end{tabular}
}
\end{table*}

\subsubsection{Influence de la masse de la primaire}

Nous avons essayé de voir comment évoluait ce modèle en faisant 
varier la masse $M_1$ attribuée à la primaire, et nous avons constaté 
que:
\begin{itemize}
\item[(1)]{
pour $M_1 < 1,2$ M$_\odot$ le rapport de masses $\mu$ devient supérieur 
à l'unité, et pour $M_1 > 2,2$~M$_\odot$ la secondaire devient 
une étoile A, possibilité exclue car on verrait alors un spectre 
composite. On doit donc avoir 1,2 M$_\odot$ $< M1 <$ 2,2 M$_\odot$;
}
\item[(2)]{
compte tenu de cette limitation pour $M_1$, le type spectral de 
la secondaire se situe pratiquement dans l'intervalle F2 V --- F8 V
}
\item[(3)]{
on a $2,7 \leq \Delta m \leq 2,8 $ 
}
\end{itemize}

Par conséquent, le modèle ``synchrone'' présenté ci-dessus est 
relativement peu influencé par la valeur de la masse de la primaire.

\subsubsection{Influence de 3e corps}

Nous allons considérer deux cas nettement distincts:
\begin{itemize}
\item[(1)]{
l'orbite du 3e corps est coplanaire avec celle du système intérieur. 
On prend donc $i = 43^{\circ}$ pour l'orbite extérieure. 
Alors, $f(m) = 0,0026$ conduit, compte tenu du modèle ci-dessus, 
à $M_3 = 0,5$~M$_\odot$ pour le troisième corps, ce qui correspond 
à la masse d'une étoile de type M0 V, plus faible de quelque 8 mag. 
que l'étoile principale du système. Dans ces conditions, on peut 
considérer comme négligeables les contributions 
du troisième corps aux paramètres $M_V$ et $B - V$ du système.
}
\item[(2)]{
l'orbite du troisième corps n'est pas coplanaire.
Si on suppose une valeur de $i$ très différente pour l'orbite extérieure, 
$i = 90^{\circ}$ par exemple, on a dans ce cas $M_3 = 0,3$~M$_\odot$, 
donc une étoile M naine encore plus froide; par contre, pour de faibles 
valeurs de $i$, la valeur de la masse du troisième corps pourra être 
plus élevée: pour $i = 20^{\circ}$ nous aurons $M_3 = 1$~M$_\odot$, 
et une étoile similaire au Soleil qui restera encore invisible, 
à quatre magnitudes d'écart avec le système principal.
}
\end{itemize}

Par conséquent, dans le ``modèle synchrone'' envisagé, le troisième corps 
serait, probablement, une étoile froide de faible masse.

\section{Conclusion}

HD 191588 est, à notre connaissance, un nouveau système de type RS CVn, 
triple de surcroît.

De par ses caractéristiques: présence d'une étoile géante, période 
relativement longue de 60 jours, il s'apparente au 
``groupe à longues périodes'' décrit par Hall (1976). Dans une étude 
consacrée au système HR 4665, qui présente certains traits communs 
avec HD 191588, Bopp et al. (1979) effectuent une intéressante discussion 
sur l'implication de ce type de binaires vis-à-vis de la classe des RS CVn. 
Ces deux systèmes ont des composantes primaires de types voisins 
(K0 III pour HR 4665), des périodes proches de 60 jours, et des orbites 
circularisées; par contre HR 4665 est une binaire à raies doubles avec 
un rapport de masses proche de l'unité. Plus généralement, Bopp et al., 
après avoir rappelé les caractéristiques des groupes de Hall, et faisant 
allusion au travail observationnel de Young \& Koniges (1977), retiennent 
un point commun à tous les systèmes ``de type RS CVn'': 
la forte émission de Ca II semble en relation avec l'existence 
d'interactions par effets de marées entre les composantes. 
Pour HD 191588, en tous cas, le fait que l'orbite soit presque circularisée 
et le probable synchronisme rotation--révolution sont des signes 
tangibles de cette interaction. 

Enfin, il convient aussi de préciser que les observations du satellite 
Hipparcos ont mis en évidence la variabilité photométrique de HD 191588. 
Dans le catalogue Hipparcos (ESA, 1997), la magnitude apparente visuelle 
de cette étoile est reportée avec le niveau de variabilité 2 
(l'échelle comportant 3 niveaux), ce qui implique, selon ce catalogue, 
une variabilité comprise entre 0,06 et 0,6 mag.; cependant la variation 
ne paraît pas corrélée avec la phase orbitale. La variabilité photométrique 
est un trait courant des étoiles de type RS CVn, qu'elle soit intrinsèque 
ou, dans certains cas, due à des éclipses. En conséquence, une étude 
photométrique de ce système serait nécessaire.

\vskip 5mm
\noindent
\textsl{Remerciements.}
Nous tenons à remercier M. Mayor, directeur de l'Observatoire de Genève, 
pour les missions accordées à l'instrument CORAVEL de l'OHP, ainsi que 
S.~Udry pour le traitement des observations en VR. Les recherches 
bibliographiques ont été effectuées grâce à la base de donnée SIMBAD 
du Centre de Données Astronomiques de Strasbourg.

\end{document}